\documentclass[preprintnumbers,amsmath,amssymb,floatfix]{revtex4}
\usepackage{graphicx}
\usepackage{amsmath}
\usepackage{bm}
\usepackage{mathrsfs}
\usepackage{color}
\usepackage{slashed}
\usepackage{epsfig}
\usepackage{dcolumn}% Align table columns on decimal point

\newcommand{\bald}[1]{{\bf #1}}

\newcommand{\cur}[1]{\mathscr{ #1}}

\newcommand{\ben}{\begin{eqnarray}}
\newcommand{\een}{\end{eqnarray}}
\newcommand{\nnu}{\nonumber\\}
\newcommand{\bef}{\begin{figure}[htb]\centering}
\newcommand{\eef}{\end{figure}}

\begin{document}

\title{An operator definition and derivation of collisional energy and momentum loss \\
in relativistic plasmas}

\author{R. B. Neufeld}
\email{bryon@ema3d.com}
\address{Electro Magnetic Applications, Lakewood, CO 80226, U.S.A.}

\author{Ivan Vitev}
\email{ivitev@lanl.gov}
\address{Los Alamos National Laboratory, Theoretical Division, MS B238, Los Alamos, NM 87545, U.S.A.}

\author{Hongxi Xing}
\email{hxing@lanl.gov}
\address{Los Alamos National Laboratory, Theoretical Division, MS B238, Los Alamos, NM 87545, U.S.A.}

\date{\today}

\begin{abstract}
We present an operator definition of the collisional energy and momentum loss 
suffered by an energetic charged particle in the presence of a medium. Our approach 
uses the energy-momentum tensor of the medium to evaluate the energy and momentum 
transfer rates. We apply this formalism to an energetic lepton or quark propagating in 
thermal electron-positron or quark-gluon plasmas, respectively. 
By using two different approaches to describe the energetic charged particle, an external 
current approach and a diagrammatic approach, we show explicitly that the operator 
method reproduces the known results for collisional energy loss from 
the scattering rate formalism. We further use our results to evaluate the 
collisional energy and momentum loss for the cases of heavy 
quark propagation through a quark-gluon plasma and energetic muon propagation 
in an electron-positron  plasma produced in a high-intensity laser field.
\end{abstract}

%\pacs{12.38.Mh,25.75Ld,25.75.Bh}

\maketitle

\section{Introduction}

In recent years, the study of the  properties of the medium created in high energy nucleus-nucleus 
collisions has attracted tremendous attention from both experiment and theory. 
In unraveling these medium properties, jet quenching~\cite{Gyulassy:2003mc}, which refers to the 
suppression of the production rate of high transverse momentum ($p_T$) leading particles and jets in 
relativistic heavy ion reactions relative to a naive superposition of nucleon-nucleon 
collisions, is thought to provide valuable information about the properties of the quark-gluon plasma 
(QGP)~\cite{Gyulassy:2003mc} and cold nuclear matter (CNM)~\cite{Neufeld:2010dz,Xing:2011fb}. 
This suppression has been attributed to the energy loss of high-$p_T$ patrons due to 
interactions between the energetic jet and the medium through elastic and inelastic scattering.  
There have been multiple studies of the energy loss based on perturbative QCD, where radiative 
energy loss~\cite{Gyulassy:1993hr, BDMPS, Gyulassy:2000er,Zakharov:1997uu,Wang:2001ifa,Arnold:2002ja} 
is thought 
to dominate the leading particle and jet attenuation and the collisional energy loss 
is relatively small. Experimental data on heavy meson attenuation from both 
RHIC~\cite{Adler:2005xv,Adare:2006nq,Aggarwal:2010xp}  and the LHC~\cite{ALICE:2012ab, Abelev:2012qh},
however, suggest that radiative energy loss alone may not be sufficient to describe the magnitude
of the observed attenuation. Collisional effects, such as energy loss and hadron dissociation, 
may play a role in heavy flavor quenching~\cite{vanHees:2004gq,Wicks:2005gt,Gossiaux:2008jv,Sharma:2009hn,Uphoff:2012gb}. 
The cumulative effect of collisional energy loss is also amplified in a parton shower 
and can be studied in jet observables~\cite{Renk:2009hv,He:2011pd,Neufeld:2012df,ColemanSmith:2012vr,Dai:2012am}, 
especially for large radii  R~\cite{Huang:2013vaa}.

The first perturbative estimate of the collisional energy loss rate $dE/dx$ was made by 
Bjorken~\cite{Bjorken:1982tu}. Subsequently, Braaten and Thoma (hereafter referred to as BT) 
performed a  calculation of $dE/dx$ where the energy loss was defined as the average over 
the interaction rate $\Gamma$ of the energy transfer $\omega$ and divided by the 
velocity $u$ of the energetic parton~\cite{Braaten:1991jj}. This is expressed in a symbolic 
formula as
\ben
\frac{dE}{dx}=\frac{1}{u}\int d\Gamma \omega,
\label{eq-BT}
\een
where the energy loss can be calculated analogously to the interaction rate from either 
the scattering matrix element or the imaginary part of the energetic 
parton self-energy.

Most calculations of collisional energy loss have focused on the perspective of the energetic particle as it propagates 
in the medium and suffers losses through scattering. A different point of view, which we will emphasize here, is the 
perspective of the medium as it observes and responds to the propagating particle. This is especially useful when we
are interested in collective phenomena in electro-magnetic and strongly-interacting 
plasmas~\cite{Nishikawa:2004ug,Mrowczynski:2007hb,Wang:2013qca} 

At a fundamental level, the properties and dynamics of a medium - including the energy transfer rate of an energetic 
particle into the medium - are contained in its energy momentum tensor defined in terms of the underlying fields.  
Explicitly, the four-momentum loss $dP^{\nu}/dt$ per unit time (throughout this paper 
we use capital letters to denote four-momentum) can be related to the spatial integration of 
the individual components of the energy momentum tensor as
\ben
\label{eloss1}
\frac{dP^{\nu}}{dt}=\int d^3x \partial_{\mu}T^{\mu\nu}(X).  
\label{eq-dEdtJ}
\een
Here, $T^{\mu\nu}$ is the medium energy-momentum tensor (EMT) and summation over repeated indices is implied.
Early analysis of the medium-energy momentum tensor response to an energetic particle was based in the desire to 
understand the medium response in the form of shockwaves or Mach cones.  The quantity $\partial_{\mu}T^{\mu\nu}(X)$ - or source term -  
in Eq.~(\ref{eloss1}) above not only contains information about the collisional energy loss but also acts as a seed for 
the fluid dynamic response of the medium to a fast particle.

Many significant attempts have  been made to understand the energy 
momentum deposition (i.e. source term) profile in QCD~\cite{Neufeld:2009ep, Qin:2009uh, Li:2010ts}. 
In the strongly-coupled limit, the AdS/CFT correspondence has been used to evaluate  the stress-tensor 
within the context of linearized gravity~\cite{Friess:2006fk, Yarom:2007ni, Chesler:2007an}. 
Recently, in the weakly-interacting limit, attempts have been made to calculate from first-principles 
the energy deposition of a fast parton traversing the QGP in terms of the medium energy-momentum 
tensor $T^{\mu\nu}$~\cite{Neufeld:2010xi,Neufeld:2011yh} to leading logarithmic accuracy. They express the collisional 
energy loss from the perspective of the medium.

In this paper, with the help of  Feynman rules that have been derived from the operator 
definition of the EMT~\cite{Neufeld:2010xi}, we will compute the collisional energy transfer rate including both 
soft and hard contributions. Our analysis shows  that the energy momentum tensor provides a natural 
and powerful way to approach collisional energy and momentum deposition, 
and can be extended to many systems of physical interest.
The outline of the paper is as follows: we present the theoretical formalism for the evaluation of the 
medium response in Sec.~\ref{sec-formalism}. In Sec.~\ref{sec-calculation} we apply our formalism to an  energetic 
lepton propagating in a thermal electron-positron plasma (EPP) by using the external current approach, 
and to energetic quark traveling through a quark-gluon plasma  using a much more general diagrammatic 
approach. We show that these two approaches  reproduce the result for collisional energy loss from the 
scattering rate formalism. We further show in Sec.~\ref{sec-result} the numerical result for the collisional 
energy-momentum transfer rate in the case of experimentally relevant QED and QCD plasmas. Our summary 
is given in Sec.~\ref{sec-summary}.

\section{Formalism}
\label{sec-formalism}

In this section we present the formalism used in the paper. We focus on the medium EMT $T^{\mu\nu}$ in the presence 
of a fast parton created in the distant past.  We are particularly interested in the divergence of the EMT, or the 
source term $J^\nu$. The source term is useful because it provides a way to obtain the energy and momentum loss 
from the medium's point of view (as shown in Eq. (\ref{eloss1})) and also drives the bulk evolution of the 
medium.  We  present our results in a general integral form, which we will use in  later sections to extract a 
specific quantity, namely the collisional energy and momentum loss of an energetic lepton or quark.

At a fundamental level, the properties and dynamics of a medium are contained in its energy-momentum tensor 
defined in terms of the underlying fields.  We begin this section by introducing this important quantity with an 
eye on how we will set up the problem of evaluating it in the presence of a fast lepton.  We consider a medium of massless 
electrons and positrons with conventional field notation: fermion fields are denoted by $\psi$ and photon fields by $A$.  
The QED EMT is given by~\cite{qedemt}
\ben
\label{qemt}
T^{\mu\nu} = \frac{i}{4}\bar{\psi}\left(\gamma^\mu\,\overset{\text{\tiny$\leftrightarrow$}}{D^\nu} 
+ \gamma^\nu\,\overset{\text{\tiny$\leftrightarrow$}}{D^\mu}\right)\psi - g^{\mu\nu}\cur{L}_F,
\een
where
\ben
\cur{L}_F = \frac{i}{2}\bar{\psi}\,\overset{\text{\tiny$\leftrightarrow$}}{\slashed{D}}\,\psi \text{,   }~~~~D^\mu = \partial^\mu - i e\,A^\mu
\een
and
\ben
\bar{\psi}\,\gamma^\mu\,\overset{\text{\tiny$\leftrightarrow$}}{D^\nu} \, \psi = \bar{\psi}\,
\gamma^\mu\,\overset{\text{\tiny$\rightarrow$}}{D^\nu} \, \psi - \bar{\psi}\,\gamma^\mu\,\overset{\text{\tiny$\leftarrow$}}{D^{*\nu}} \, \psi.
\een
In the above equations, $e$ is the electromagnetic coupling parameter and conventional slashed notation is used, 
$\slashed{A} = \gamma_\mu A^\mu$, etc.  A summation over spin is implied in the EMT.  

In principle, any quantity relating to the energy and momentum of a medium can be extracted from the EMT.  For example, 
one can use the EMT in thermal field theory to obtain perturbative corrections to the pressure or energy density of an ideal gas.  
Furthermore, one does not need a system in equilibrium to use the EMT.  Any distribution function, whether or not it is thermal, 
can be used with the EMT to extract quantities of interest. The utility and breadth of applications of the EMT make 
it a powerful analytical tool to investigate medium properties, and provide one possible approach to bridge the short and 
long distance dynamics of a medium.  

In order to calculate the medium response far from the fast parton, one can consider two possibilities in terms of EMT: 
a) calculate individual component of the EMT directly within perturbation theory to obtain information about the medium response; 
b) calculate the source term for the EMT within perturbation theory and use an effective theory to propagate the resulting 
disturbance to regions far from the fast parton. In some sense, the two approaches are related since both of them arise from the 
EMT. However, there is essential difference between these two: the additional derivatives in the source term serve to 
add momentum weighting, this additional weighting  changes the infrared behavior completely and make the final result 
infrared safe. Therefore, we apply approach (b) in this paper to consider the problem of collisional energy and momentum 
loss in a thermal medium, or medium response, so that we can get a physical result that independent of any cut-off scales. 
This application allows us to obtain analytical results that compare directly to previous results obtained using scattering rates. 
The rate of energy and momentum being transferred to a medium is related to the EMT through the equations
\ben
\label{eloss2}
\frac{dE}{dt}&=&\int d^3x  \partial_{\mu}T^{\mu 0}(X),\nnu
\frac{dp^i}{dt}&=&\int d^3x  \partial_{\mu}T^{\mu i}(X).
\een
In an isolated medium, Eq.~(\ref{eloss2}) of course evaluates to zero because of energy momentum conservation.  However, 
when a fast projectile is being pushed through the surrounding medium, which can be represented by some external current, 
it must be transferring energy and momentum to the surrounding medium,  meaning  that the  EMT is not conserved. 
Assuming we have some way to couple an external source of energy to the medium, the evaluation of the energy transfer 
rate can be performed using standard techniques of thermal field theory. In the rest of this section we will present the 
basic details of such an evaluation and present a formula that can be used for the calculation of energy momentum transfer 
rates to a medium for a wide variety of problems.

\bef
\psfig{file=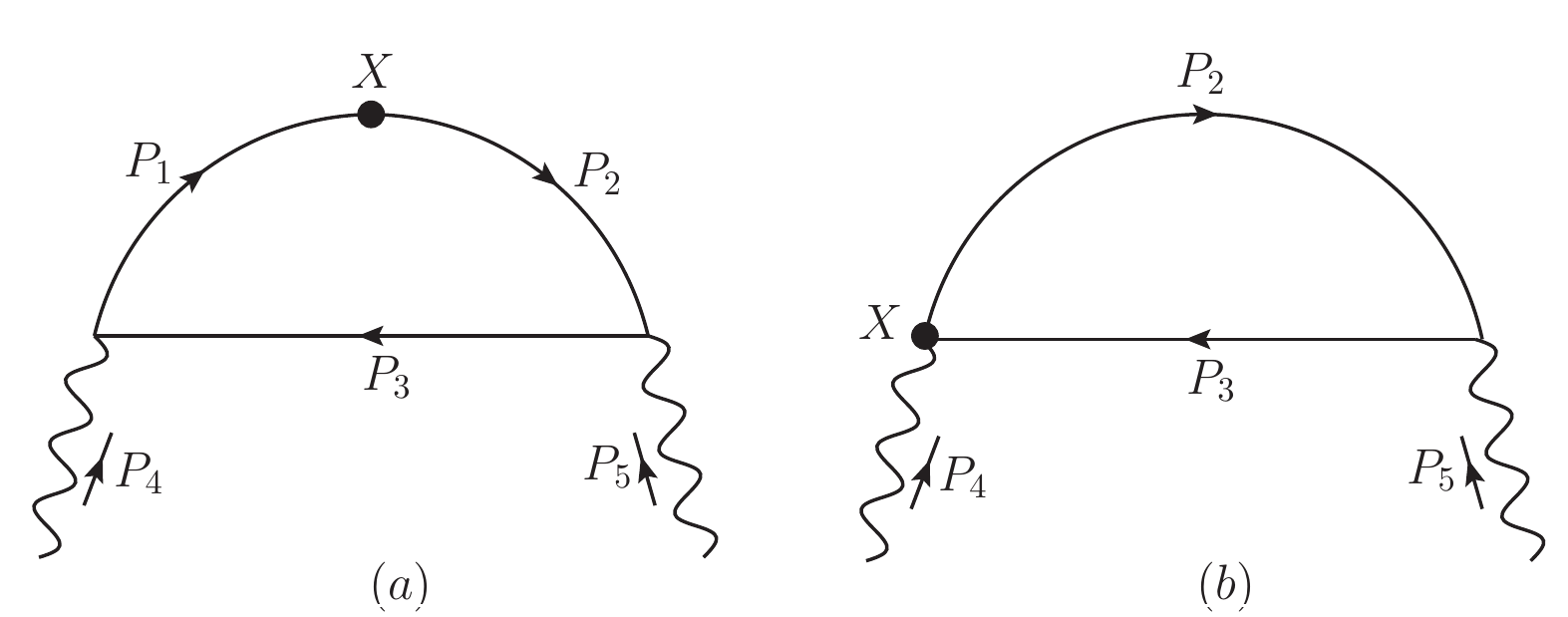, width=3.5in}
\caption{Feynman diagrams contributing to $\langle \partial_\mu T^{\mu\nu}(x)\rangle$ in the presence of a source 
interaction term, $A_\mu^a \, j^\mu_a$.   The diagram in Figure 1(a) can be traced back to terms in the energy-momentum tensor 
(see equation (\ref{qemt})) which go as $\bar{\psi}\gamma\partial\psi$, whereas the diagram in Figure 1(b) originates 
from terms that go as $g\,\bar{\psi}\gamma A\psi$.
}
\label{feyn1}
\eef
We will borrow some of the results presented in~\cite{Neufeld:2010xi}, where the EMT source term was calculated for 
a medium in the presence of a classical charge in the  hard thermal loop (HTL) approximation. In the HTL limit
the fields generated by the external current are soft compared to the medium temperature. Explicitly, the fermion propagator 
is expanded in the limit where the momentum of the exchanged gluon is  much smaller than the one of the medium parton.

Building upon  the analysis presented in the  previous work, the presentation here will be extended to a much more general 
case by including both the hard and soft contributions.  The technique of introducing an arbitrary momentum scale 
$q^*$ to separate the hard  and soft regions of the momentum transfer was developed in~\cite{Braaten:1991jj}. 
The contribution from hard momentum transfers is computed by using 
the tree-level propagator for the exchanged gluon, while the contribution from the soft region is computed using 
an effective gluon propagator. The dependence on the arbitrary scale $q^*$ cancels upon adding the hard and 
soft contributions.

The lowest order Feynman diagrams for energy transfer into an electron-positron plasma is shown in Fig.~\ref{feyn1}.  
We do not have to consider yet what the external photons connect to, only that they will represent a two photon exchange 
with the medium.  The next section will provide a specific application toward collisional energy and momentum loss, 
but in this section the source of energy remains general. That the  lowest order diagrams for energy-momentum exchange 
requires two photon exchange can be easily verified using Furry's theorem. The same is true for QCD. Since we are focused 
on the energy-momentum transfer rate in this paper, the diagrams of Fig.~\ref{feyn1} should be evaluated using real time thermal 
field theory, where the diagram in Fig.~\ref{feyn1}(a) comes from the bare part of the EMT (the pieces with no coupling constant), 
and Fig. \ref{feyn1}(b) arises from the interacting part of the EMT.  The interested reader can easily verify this noting that 
the bare part of the EMT contains two fields each at position $X$, while the interacting contains three fields at $X$.

The diagrams of Fig.~\ref{feyn1} can be evaluated using standard Feynman rules (for instance Das~\cite{Das:1997gg}). However, 
we must also consider the effect of the derivative structure from taking the divergence of the EMT. The derivatives 
serve to add momentum weighting to what the diagrams by themselves would yield.  Specifically, we are interested in 
evaluating the unique momentum contributions from the term  $\partial_{\mu}T^{\mu \nu}(x)$ in Eq.~(\ref{eloss2}). Using the 
momentum convention shown in Fig.~\ref{feyn1}(a), we find that taking the divergence of the EMT yields a momentum weighting of 
\ben
\label{fig2a}
\frac{i e^{-i X \cdot(P_1 - P_2)}}{4} \times\left[(P_2^2 - P_1^2) \gamma^{\nu}  
+ P_1^{\nu} (3 \slashed{P}_2 + \slashed{P}_1) -  P_2^{\nu}(3\slashed{P}_1 + \slashed{P}_2) \right].
\een
The exponential term arises since the EMT is evaluated in position space.  

For the diagram in Fig.~\ref{feyn1}(b) we choose the convention that the photon momentum flows into the interaction position $X$.  
Therefore, the EMT contribution is
\ben
\label{fig2b}
-i e \,e^{-iX\cdot(P_4 + P_3 - P_2)}\,(P_4 + P_3 - P_2)_\mu \times\frac{\left(\gamma^\nu g^{\mu\sigma} 
+ \gamma^\mu g^{\nu\sigma} - 2 g^{\mu\nu}\gamma^\sigma\right)}{2}.
\een
Note that Eqs.~(\ref{fig2a}) and~(\ref{fig2b}) are only the contribution from taking the divergence of the EMT 
in the diagrams of Fig.~\ref{feyn1}.  It is still necessary to evaluate the diagrams in a  conventional manner 
to get the rest of the contribution. The full result for Fig.~\ref{feyn1} is obtained using Feynman rules at finite temperature, 
and we will go through a few of the steps.  
Starting with Fig.~\ref{feyn1}(a) and using the results from Eq. (\ref{fig2a}) we have:
\ben
\label{barefirst}
\partial_\mu {T}_0^{\mu\nu}(X) &=& -\frac{e^2}{4} \int \frac{d^4 P_i}{(2\pi)^{12}} e^{-i X \cdot(P_1 - P_2)} 
\text{Tr}\left[\left((P_2^2 - P_1^2) \gamma^{\nu}  + 3 \slashed{P}_2 P_1^{\nu}  + \slashed{P}_1 P_1^{\nu} 
 -  \slashed{P}_2 P_2^{\nu}  - 3\slashed{P}_1 P_2^{\nu}  \right)\slashed{P}_1\gamma^\sigma\slashed{P}_3\gamma^\omega\slashed{P}_2\right]\nnu
&&\times\left[\left(T(P_3)G_R(P_1) + T(P_1)G_A(P_3)\right)G_A(P_2)+T(P_2)G_R(P_3)G_R(P_1)\right]\nnu 
&&\times D_{\sigma \tau}(P_4)D_{\omega \lambda}(P_5)\delta^4(P_2 + P_5 - P_3)\delta^4(P_1 - P_4 - P_3)  \otimes F^{\tau\lambda}(P_4,P_5) ,
\een
where the notation ${T}_0^{\mu \nu}(X)$ means this is the contribution from the bare (or without coupling constant) part of the EMT.
A few comments are in order regarding Eq.~(\ref{barefirst}).  First, the notation $\int d^4 P_i$ means that 
all momenta are integrated over $\int d^4 P_i=\int d^4 P_1 d^4 P_2 d^4 P_3 d^4 P_4 d^4 P_5$.   
$D_{\sigma\tau}(P)$ is the propagator for the exchanged photon.   In the limit of a hard momentum exchange $D_{\sigma\tau}(P)=(-g_{\sigma\tau})G_R(P)$, 
however in the soft region one must use an HTL resummed propagator for the photon exchange.  More will be said on this below.
The Green's function notation is
\ben
G_{R/A}(P) = \frac{1}{P^2 \pm i\epsilon P^0}
\label{eq-green}
\een
and $T(P)$ is the medium's particle distribution function.  For a thermal system of massless fermions, we have
\ben
T(P) = 2\pi i n_F(|P^0|)\delta(P^2).
\een
However, as pointed out above, there is no reason one has to use a thermal medium.
Finally, the notation $\otimes F_{\tau\lambda}(P_4,P_5)$ simply indicates that the expression above contains explicitly only 
the contribution from the diagrams related to the EMT.  One must attach the photons in Fig.~\ref{feyn1} to some external source 
of energy-momentum to obtain a non-zero result.

We can perform the same analysis on the diagram in Fig.~\ref{feyn1}(b)
\ben
\label{intfirst}
\partial_\mu {T}_I^{\mu \nu}(X) &=&  -\frac{e^2}{2}\int \frac{d^4 P_i}{(2\pi)^{12}} 
e^{-i X \cdot(P_4 + P_5)} \text{Tr}\left[\left(\gamma^{\nu}\,g^{\mu \sigma} + \gamma^\mu\,g^{\nu\sigma} 
- 2 g^{\mu\nu}\gamma^\sigma\right)\slashed{P}_2\gamma^\omega\slashed{P}_3\right](P_4 + P_3 - P_2)_\mu \nnu
&&\times  D_{\sigma \tau}(P_4)D_{\omega \lambda}(P_5)\left[T(P_2)G_R(P_3) + T(P_3)G_A(P_2) \right]
\delta^4(P_2 + P_5 - P_3)\otimes F^{\tau\lambda}(P_4,P_5) ,
\een
where the notation ${T}_I^{\mu \nu}(X)$ means this is the contribution from the interacting 
(with coupling constant) part of the EMT.

What remains to be done is to combine and simplify Eqs.~(\ref{barefirst}) and~(\ref{intfirst}).  
Since we are specifying a thermal 
medium, we have used the relation that $P^2 T(P) = 0$.  We have also made use of simplifications such as $P^2 G_R(P) = 1$ 
and enforced the $\delta$-functions constraints. The final result is 
\ben
\label{emtpiece}
\partial_\mu {T}^{\mu \nu}(X) &=& 8\,e^2 \int \frac{d^4 P_i}{(2\pi)^{12}} e^{-i X \cdot(P_4 + P_5)} T(P_3) 
D_{\sigma \tau}(P_4)D_{\omega \lambda}(P_5) G_A(P_3 - P_4)\nnu
&&\times\left[2\,({P_3}^\sigma\,{P_3}^\omega\,- {P_3}^\sigma\,{P_4}^\omega\,)P_5^{\nu} + g^{\sigma\omega} \,
P_3\cdot P_4\,P_5^{\nu} - g^{\nu\sigma}\,{P_3}^\omega(2 P_3\cdot P_5 - P_4\cdot P_5)\right] \otimes F^{\tau\lambda}(P_4,P_5).
\een
To obtain the energy transfer rate will require an integration over all space, as indicated in
 Eq.~(\ref{eloss2}). Eq.~(\ref{emtpiece}) contains 
a fairly general expression for the energy momentum transfer rate to an electromagnetic plasma.  
It was obtained from the divergence 
of the EMT without specifying  the source of energy and momentum, except that it be coupled via  
two photon exchange.  However, to obtain a closed-form 
energy transfer rate we must specify $F^{\tau\lambda}(P_4,P_5)$ in Eq.~(\ref{emtpiece}).  
This will be done for the case 
of collisional energy loss in the next section.

\section{Details of the Calculation  and Analytic Results}
\label{sec-calculation}

As discussed throughout the paper, our goal is to evaluate the energy-momentum transfer rate, or 
collisional energy and momentum loss, 
of an energetic particle propagating through an EPP or a QGP.  
One possibility is to add an interaction 
term to the Lagrangian of the form $\cur{L} \rightarrow \cur{L} - A_\mu \, j^\mu$, 
which was presented in Ref.~\cite{Neufeld:2010xi}.  
Here, $j^\mu$ is a classical charged current of the form $j^\mu = e U^\mu  \delta(\bald{x} - \bald{u} t)$, $U^\mu = (1,\bald{u})$,
which represents the propagating particle. We will apply this external current approach  in the Sec.~\ref{subsec-QED}.
The more general approach, which will be used in Sec.~\ref{subsec-QCD}, is to treat the energetic quark as a field, rather 
than a classical current. In this diagrammatic approach the field interacts with the medium through a two-gluon exchange. 

\subsection{Collisional energy loss in a QED plasma}
\label{subsec-QED}
\subsubsection{External current approach}
Studies of parton energy loss in the QGP are of great phenomenological interest to  heavy-ion physics. It is, however, also 
instructive to discuss the problem of energy loss in QED. In this case, it is natural to consider an asymptotic particle 
traveling  through a domain containing a relativistic electron-positron plasma. Following the approach 
developed in~\cite{Neufeld:2010xi}, 
we model the asymptotically fast lepton by an external current $j^{\mu}(X)$, which in turn can be expressed in momentum space as
\ben
j^{\mu}(K)=-2\pi ie U^{\mu}\delta(K\cdot U).
\een
Here, $K$ is the four-momentum exchanged with the medium, and $U$ is the four-velocity of the propagating fast lepton. 
\bef
\psfig{file=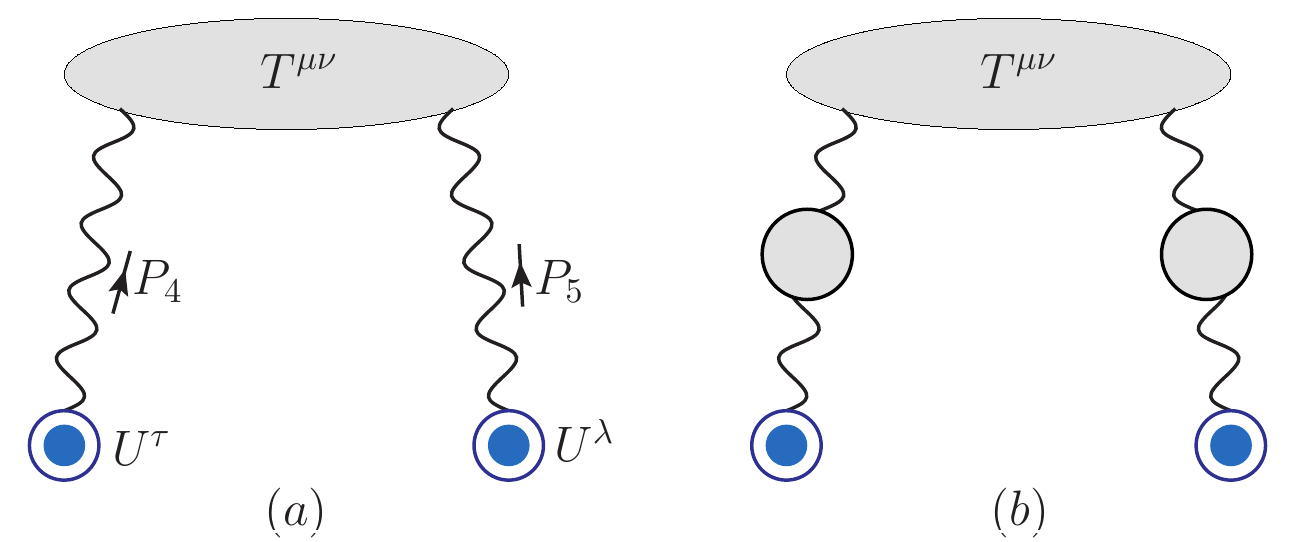, width=3.5in}
\caption{Feynman diagrams for the source term in QED in the presence of the external current $J^{\tau}\sim U^{\tau}$, 
which is represented by the blue circle. The upper gray loop represents the EMT contribution, as shown in Fig. \ref{feyn1}. 
Diagram (a) is the tree level diagram, which contributes to the hard region. Diagram (b) contributes to the soft 
region and the hard thermal loop resummed photon propagator is represented by a gray blob.}
\label{fig-QED}
\eef

In Fig. \ref{fig-QED} we focus  on the contribution from $F^{\tau\lambda}(P_4,P_5)$ to Eq. (\ref{emtpiece}).  
The upper loop, which represents the EMT contribution, is shown for completeness and has been addressed in the previous section.
All that is left to do is to use Feynman rules to evaluate the contribution from the fast lepton to the energy transfer rate.  
We find it contributes
\ben
\label{sourcecont1}
F^{\tau\lambda}(P_4,P_5)= -(2\pi)^2e^2U^{\tau}U^{\lambda}\delta(P_4\cdot U)\delta(P_5\cdot U).
\een
Combining Eq. (\ref{sourcecont1}) with Eq. (\ref{emtpiece}) gives 
\ben
\label{finalemt1}
\partial_\mu {T}^{\mu \nu}(X) &=& -8e^4 \int \frac{d^4 P_i}{(2\pi)^{10}} e^{-i X \cdot(P_4 + P_5)} T(P_3) 
D_{\sigma \tau}(P_4)D_{\omega \lambda}(P_5) G_A(P_3 - P_4)U^{\tau}U^{\lambda}\delta(P_4\cdot U)\delta(P_5\cdot U)\nnu
&&\times\left[2\,({P_3}^\sigma\,{P_3}^\omega\,- {P_3}^\sigma\,{P_4}^\omega\,)P_5^{\nu} + g^{\sigma\omega} \,
P_3\cdot P_4\,P_5^{\nu} - g^{\nu\sigma}\,{P_3}^\omega(2 P_3\cdot P_5 - P_4\cdot P_5)\right] ,
\een
The result of Eq.~(\ref{finalemt1}) can then be inserted in Eq. (\ref{eloss2}) to obtain the  collisional energy-momentum  loss.  
In the next subsection we will evaluate  the collisional energy loss and compare our results to the ones 
obtained using conventional scattering methods.

Before we undertake the evaluation of  Eq. (\ref{finalemt1}), a few words on soft and hard contributions  
at finite temperature are in order.  Calculations at finite temperature involving soft excitations have been known 
for some time to require resummation techniques to obtain gauge-invariant results~\cite{Braaten:1989kk}.  On the other hand, 
calculations involving hard excitations do not require resummation techniques and one can apply bare perturbation theory. 
Soft is here formally defined as a quantity of order $eT$ and hard is a quantity of order 
$T$ or larger, where $T$ is the temperature and $e$ the coupling constant with $e \ll 1$.  The contributions 
from the hard and soft excitations should be matched consistently. A general method for carrying out the matching 
at finite temperature is to separate the soft and hard regimes by a separation parameter $e T \ll q^* \ll T$~\cite{Braaten:1991dd}, 
and then use whatever technique is most efficient in each regime separately.  The final result is obtained by 
combining the two separate regimes into one complete calculation.  If the final result is independent of $q^*$ and 
the result of each regime is independently gauge invariant.

We will employ this type of separation in our calculation, and verify explicitly that the final result is independent 
of the details of the separation.  In practice, this means that the photon propagators of Eq.~(\ref{eq-green}) must include 
the HTL resummation when evaluating the soft contribution.  For the hard contribution, no such resummation is necessary.

\subsubsection{Hard contribution}

Starting with the expression  for collisional energy loss in Eqs. (\ref{finalemt1}) and (\ref{eloss2}), the 
contribution from the hard momentum exchange region can be cast into
\ben
\left.\frac{dE}{dt}\right|_{hard}&=&8e^4\int\frac{d^4P_3d^3p_4}{(2\pi)^{7}}T_F(P_3)G_A(P_3-P_4)[G_R(P_4)]^2
{\bf p}_4\cdot {\bf u}\left[P_3\cdot P_4U^2+2(P_3\cdot U)^2\right],
\label{eq-dEdt-h1}
\een
where we  have eliminated the terms associated with $P_4^2-2P_3\cdot P_4$ because these terms cancels with $G_A(P_3-P_4)$ and the resulting $P_3$ 
integration vanishes by symmetry. In Eq. (\ref{eq-dEdt-h1}), $P_3^0$ can be integrated out by taking advantage of the $\delta$-function 
in $T_F$, then the hard contribution reduces to
\ben
\left.\frac{dE}{dt}\right|_{hard}=-4e^4\int\frac{d^3{p_3}d^3{p_4}}{(2\pi)^{5}p_3}n_F(p_3)
[G_R(P_4)]^2{{\bf p}_4}\cdot{{\bf u}}\left[2(P_3\cdot U)^2+U^2P_3\cdot P_4\right]\delta(P_4^2-2P_3\cdot P_4){\rm sgn}(p_3-{\bf p}_4\cdot {\bf u}).
\een
In order to simplify the expression further, it is convenient to make use of the fact that in a static medium 
(we can always work in its local
rest frame) the collisional energy loss does not depend on the direction of ${\bf u}$. We can, therefore, specify a coordinate 
system with ${\bf u}$ in the ${\bf z}$ direction  evaluate the angular integrals 
in $\int d^3p_3$ and $\int d^3p_4$. On the other hand, the kinematics of the interaction between the fast lepton and the 
medium constrain the integration limit of $p_4$ and $\nu$ as follows
\ben
p_4<\frac{2p_3}{1+u\nu}, ~~~\nu<\frac{2p_3-p_4}{p_4u},
\een
where $\nu={\bf u}\cdot \hat{\bf p}_4$ denotes the angle between the incident fast lepton and the exchanged photon. 
The kinematics limits 
will make ${\rm sgn}(p_3-{\bf p}_4\cdot {\bf u})$ always positive. The collisional energy loss for the hard momentum exchange reduces to 
\ben
\left.\frac{dE}{dt}\right|_{hard}&=&-e^4\int\frac{dp_3}{(2\pi)^{3}}
\left[\int_{q^*}^{\frac{2p_3}{1+u}}\frac{dp_4}{p_4^2}\int_{-1}^1d\nu+\int_{\frac{2p_3}{1+u}}^{\frac{2p_3}{1-u}}
\frac{dp_4}{p_4^2}\int_{-1}^{\frac{2p_3-p_4}{p_4u}}d\nu\right]\frac{n_F(p_3)}{(1-u^2{\nu}^2)^2}\nnu
&&\times\left[2p_3^2\left(2-4 u\nu\omega+u^2(1-\omega^2)+u^2\nu^2(3\omega^2-1)\right)-(1-u^2)p_4^2(1-u^2{\nu}^2)\right]u{\nu},
\label{eq-dEdt-h2}
\een
where
\ben
\omega=\frac{p_4(1-u^2\nu^2)+2p_3u\nu}{2p_3}.
\een
The maximum momentum transfer, providing the upper limit for the integration in the above equation, 
has been determined from the kinematics of 
the scattering (for discussion of kinematic limits effects see~\cite{Ovanesyan:2011xy}).

We have enforced a lower limit for the $\int dp_4$ integration to regularize the infrared divergence, so that the remaining 
integrals can be evaluated analytically. Therefore, the final result of the hard contribution is free of infrared divergences. 
However, it depends on this cut-off scale $q^*$ and the dependence is included in the leading logarithmic term
\ben
\left.\frac{dE}{dt}\right|_{hard}=\frac{e^4T^2}{24\pi}\left[1-\frac{1-u^2}{u}tanh^{-1}[u]\right]
\left(\ln\frac{T}{q^*}+\ln\frac{1}{\sqrt{1-u^2}}+C_h(u)\right)
\label{eq-dEdt-h},
\een
where $C_h(u)$ is the constant term. As we can see from the above result, the logarithmic infrared divergences in the tree 
level diagrams manifest themselves as logarithms of $q^*$.  This behavior arises from  long range interactions mediated by the photon. 
In principle, this long range interactions should be screened in the medium, therefore, one need to resum the hard thermal 
loop corrections which take into account the screening.

\subsubsection{Soft contribution}
The soft contribution to collisional energy loss has been calculated in imaginary-time formalism in terms of the imaginary part 
of the self-energy of the projectile lepton. Here, we perform the calculation in real time formalism for thermal field theory 
in terms of the energy-momentum tensor. In the region of phase space where the exchanged photon is soft, hard thermal loop 
corrections to the photon propagator must be resummed, the net effect is to replace the bare photon propagator as $D^{\mu\nu}(Q)$. 
In the Coulomb gauge it is given by
\ben
D^{\mu\nu}(Q)=-P_L^{\mu\nu}\Delta_L(q_0,q)-P_T^{\mu\nu}\Delta_T(q_0,q),
\label{eq-propagator}
\een
where the longitudinal projector $P_L^{\mu\nu}=\delta^{\mu 0}\delta^{\nu 0}$, and the transverse projector $P_T^{00}=0$, 
$P_T^{ij}=\delta^{ij}-\hat q^i \hat q^j$. The effective longitudinal and transverse propagators are
\ben
\Delta_L^{-1}(q_0,q)&=&q^2-\frac{3}{2}m_{\gamma}^2\left[\frac{q_0}{q}\ln\frac{q_0+q}{q_0-q}-2\right],\nnu
\Delta_T^{-1}(q_0,q)&=&q_0^2-q^2+\frac{3}{2}m_{\gamma}^2\left[\frac{q_0(q_0^2-q^2)}{2q^3}\ln\frac{q_0+q}{q_0-q}-\frac{q_0^2}{q^2}\right],
\een
where $m_{\gamma}=eT/3$ is the photon screening mass. Inserting this effective photon propagator into the expression for collisional 
energy loss, the nonzero contribution from the Dirac traces is  
\ben
\left.\frac{dE}{dt}\right|_{soft}=-8e^4\int\frac{d^4P_3d^3p_4}{(2\pi)^7}T_F(P_3)G_A(P_3-P_4) 
\left[\left|\Delta_L(P_4)\right|^2H_{LL}+2Re\left(\Delta_L(P_4)\Delta_T^*(P_4)\right)H_{LT}
+\left|\Delta_T(P_4)\right|^2H_{TT}\right]. \quad
\een
Here, $H_{LL}$ and $H_{TT}$ arise from the longitudinal and transverse components of the effective photon 
propagator,  respectively, and $H_{LT}$ is the interference between them
\ben
H_{LL}&=&{\bf p}_4\cdot{\bf u}\left[P_3\cdot P_4-2p_3^0{\bf p}_4\cdot{\bf u}+2(p_3^0)^2\right],\nnu
H_{LT}&=&{\bf p}_4\cdot{\bf u}\left[\hat{\bf p}_4\cdot{\bf u}
\left({\bf p}_3\cdot\hat{\bf p}_4{\bf p}_4\cdot{\bf u}-2p_3^0{\bf p}_3\cdot\hat{\bf p}_4\right)-{\bf p}_4\cdot{\bf u}
\left({\bf p}_3\cdot{\bf u}+p_3^0\right)+2p_3^0{\bf p}_3\cdot{\bf u}\right],\nnu
H_{TT}&=&{\bf p}_4\cdot{\bf u}\left[-2\left({\bf p}_3\cdot{\bf u}-{\bf p}_3\cdot\hat{\bf p}_4\hat{\bf p}_4\cdot{\bf u}\right)^2
+P_3\cdot P_4\left((\hat{\bf p}_4\cdot{\bf u})^2-u^2\right)\right].
\een
We further integrate over $p_3^0$ by taking advantage of the delta function $\delta(P_3^2)$ in $T_F(P_3)$. In the soft region, 
where $p_4^0\ll p_3$ and $p_4\ll p_3$, the $\delta$ function $\delta(P_4^2-2P_3\cdot P_4)$, which arises 
from the imaginary part of $G_A(P_3-P_4)$, 
reduces to  $\delta(\omega-\hat{\bf p}_4\cdot{\bf u})$, with $\omega=\hat{\bf p}_3\cdot\hat{\bf p}_4$. Therefore, 
the soft contribution to the energy loss reduces to
\ben
\left.\frac{dE}{dt}\right|_{soft}&=&e^4\int\frac{d^3p_3d^3p_4}{(2\pi)^5p_3^2p_4}n_F(p_3)
\delta(\omega-\hat{\bf p}_4\cdot{\bf u}){\rm sgn}(p_3-{\bf p}_4\cdot{\bf u})\nnu
&&\times\left[\left|\Delta_L(P_4)\right|^2H_{LL}+2Re\left(\Delta_L(P_4)\Delta_T^*(P_4)\right)H_{LT}
+\left|\Delta_T(P_4)\right|^2H_{TT}\right].
\een

In a statical medium, the collisional energy loss does not depend on the direction of ${\bf u}$. 
Therefore, $dE/dt$ can be further simplified by averaging the integrand over the direction of ${\bf u}$ by using 
the following formulas
\ben
&&\int\frac{d\Omega}{4\pi}\delta(\omega-\hat{\bf p}_4\cdot{\bf u})=\frac{1}{2u}\theta(u^2-\omega^2),\nnu
&&\int\frac{d\Omega}{4\pi}\delta(\omega-\hat{\bf p}_4\cdot{\bf u})u^i=\frac{1}{2u}\theta(u^2-\omega^2)\omega\hat{\bf p}_4^i,\nnu
&&\int\frac{d\Omega}{4\pi}\delta(\omega-\hat{\bf p}_4\cdot{\bf u})u^iu^j
=\frac{1}{2u}\theta(u^2-\omega^2)\frac{1}{2}\left[(u^2-\omega^2)\delta^{ij}+(3\omega^2-u^2)\hat{\bf p}_4^i\hat{\bf p}_4^j\right],
\een
where $\int d\Omega$ represents integration over the angles of ${\bf u}$. Because of the $\theta$-function in the above angular 
integration of ${\bf u}$, the integration limits of $p_4^0$ are constrained to the space-like 
interval $-up_4<p_4^0<up_4$, these constraints 
make ${\rm sgn}(p_3-{\bf p}_4\cdot{\bf u})$ positive. On the other hand, we enforce an arbitrary upper 
limit cutoff $q^*$ (but the same as that in the hard region) 
to the integration of $\int dp_4$, thus the soft contribution to energy loss reduces to
\ben
\left.\frac{dE}{dt}\right|_{soft}&=&\frac{2}{u}\frac{e^4}{(2\pi)^3}\int dp_3p_3n_F(p_3)\int_0^{q^*}dp_4
\int_{-up_4}^{up_4}dp_4^0(p_4^0)^2\nnu
&&\times\left[|\Delta_L(P_4)|^2+\frac{1}{2}\left(1-\left(\frac{p_4^0}{p_4^2}\right)^2\right)
\left(u^2-\left(\frac{p_4^0}{p_4^2}\right)^2\right)
|\Delta_T(P_4)|^2\right].
\een
This result matches that by BT~\cite{Braaten:1991jj}, and it can be further simplified by performing 
the remaining integrations of $\int dp_4$ and $\int dp_4^0$, the integrals can be evaluated analytically 
up to the leading logarithmic term,
\ben
\left.\frac{dE}{dt}\right|_{soft}&=&\frac{e^4T^2}{24\pi}\left[1-\frac{1-u^2}{u}tanh^{-1}[u]\right]\left(\ln\frac{q^*}{3m_{\gamma}}+C_s(u)\right),
\label{eq-dEdt-s}
\een
where the constant term $C_s(u)$ can be evaluated numerically. The expression for it can 
be found in Eq.~(41) of~\cite{Braaten:1991jj}. 
Notice that the dependence on the arbitrary scale $q^*$ only exist in the leading logarithmic term. As  anticipated, 
it exactly cancels that from the hard contribution in Eq.~(\ref{eq-dEdt-h}). 

\subsubsection{Complete result}
The complete result for the collisional energy loss to leading order is the sum of the hard contribution in Eq.~(\ref{eq-dEdt-h}) 
and soft contribution in Eq.~(\ref{eq-dEdt-s})
\ben
\frac{dE}{dt}&=&\frac{e^4T^2}{24\pi}\left[1-\frac{1-u^2}{u}tanh^{-1}[u]\right]
\left(\ln\frac{T}{3m_{\gamma}}+\ln\frac{1}{\sqrt{1-u^2}}+C(u)\right),
\label{eq-dEdt-t}
\een
where the constant term $C(u)=C_h(u)+C_s(u)$. As we can see from the final result, the dependence on the arbitrary scale $q^*$ that separates 
the hard and soft regions of the momentum transfer $p_4$ cancels, leaving a logarithm of $1/e$. Importantly, by comparing our result for the 
collisional energy loss to the one in Ref.~\cite{Braaten:1991jj}, we see that the collisional energy loss from the EMT formalism 
and the scattering rate formalism are exactly the same for both the leading logarithmic term and the constant term.  There  is  
an overall minus sign difference, which indicates that the energy lost by the charged fermion is transferred to the medium completely.

Similarly, one can  obtain the collisional momentum loss by substituting the source term from Eq.~(\ref{eloss2}) into the definition of 
momentum loss in Eq.~(\ref{finalemt1}).  We found that the collisional momentum loss is closely related to energy loss
\ben
\frac{dp^z}{dt}=\frac{1}{u}\frac{dE}{dt},
\label{eq-dpdt}
\een
here we have chosen the fast lepton to propagate in $z$-direction. In this case, the linear momenta in $x$- and $y$-directions 
are conserved: $dp^x/dt=dp^y/dt=0$.

In the ultrarelativistic limit $u\to 1$, Eq. (\ref{eq-dEdt-h1}) for the hard contribution to energy loss  breaks down since the upper limit 
of the momentum transfer $p_4$ goes to infinity. In this case, one must enforce an upper limit $q_m$ on the momentum transfer,  
use it in the   other part of the calculation, and  take the $u\to 1$ limit. Thus, the hard contribution can be written as 
\ben
\left.\frac{dE}{dt}\right|_{hard}^{u\to 1}=\frac{e^4T^2}{48\pi}\left[\ln\frac{q_mT}{(q^*)^2}+\frac{8}{3}-12\ln(A)+\ln(4\pi)\right],
\een
where $A$ is Glaisher's constant with numerical value $A\simeq 1.282$.
The soft contribution is simply the $u\to 1$ limit of Eq. (\ref{eq-dEdt-s}), which is the same as that from the 
scattering rate (Eq. (62) in Ref. \cite{Braaten:1991jj}):
\ben
\left.\frac{dE}{dt}\right|_{soft}^{u\to1}&=&\frac{e^4T^2}{24\pi}\left[\ln\frac{q^*}{3m_{\gamma}}+0.256\right].
\een
Adding the hard and soft contributions together, the dependence on the separation scale $q^*$ cancels and the total collisional energy loss 
rate is 
\ben
\left.\frac{dE}{dt}\right|^{u\to 1}=\frac{e^4T^2}{48\pi}\left[\ln\frac{E}{e^2T}+2.725\right].
\een

For the sake of completeness, we mention that the operator (source term) definition of collisional energy loss
can also be implemented using kinetic theory, but only for the soft contribution.  Hence,  to obtain a full result
(both soft and hard) the field theory approach is necessary.  The kinetic theory implementation to extract the EMT coupled 
to a source of energy can be found in Ref.~\cite{Neufeld:2008hs}, 
where a QCD plasma was considered. It uses a Vlasov equation in which the external force is generated by a classical charged 
current. The EMT is obtained by taking momentum moments of the resulting Vlasov equation and one can obtain the energy transfer rate in the 
same manner as suggested in Eq.~(\ref{eloss2}) of this paper. The kinetic theory approach is more straightforward, 
but also less versatile.  It cannot be used to extract the hard contribution, and it is unclear how one can apply a more 
general external sources of energy, such as a parton shower.

\subsection{Collisional energy loss in the QGP}
\label{subsec-QCD}
The quark energy loss in a QGP is readily derived from the QED result, the contribution from the quark component of the 
medium is obtained by multiplying Eq. (\ref{eq-dEdt-t}) 
by the number of active quark flavors $N_F$ and by the color factor $C_2=(N_c^2-1)/(4N_c)$. Introducing the QCD coupling 
by $e^2\to g^2$, we obtain the quark collisional energy loss in QGP 
\ben
\frac{dE}{dt}&=&\frac{g^4T^2C_2N_F}{24\pi}\left[1-\frac{1-u^2}{u}tanh^{-1}[u]\right]\left(\ln\frac{T}{3m_{g}}+\ln\frac{1}{\sqrt{1-u^2}}+\mathcal{O}(1)\right),
\label{dEdt-QCD}
\een
where $m_g=\frac{gT}{\sqrt{3}}(1+N_F/6)^{1/2}$ is the gluon Debye mass in the QGP. Let us clarify what the above result means. 
For direct comparison to the
electron-positron plasmas we have shown only the quark-quark scattering channel. The HTL propagator, however, does 
include gluon fluctuations as can be seen from the
expression for $m_g$. Note, that if we include quark-gluon scattering the leading logarithmic result is obtained as
$ N_F/6 \rightarrow 1+  N_F/6 $. Here, the $t$-channel scattering dominates.

\bef
\psfig{file=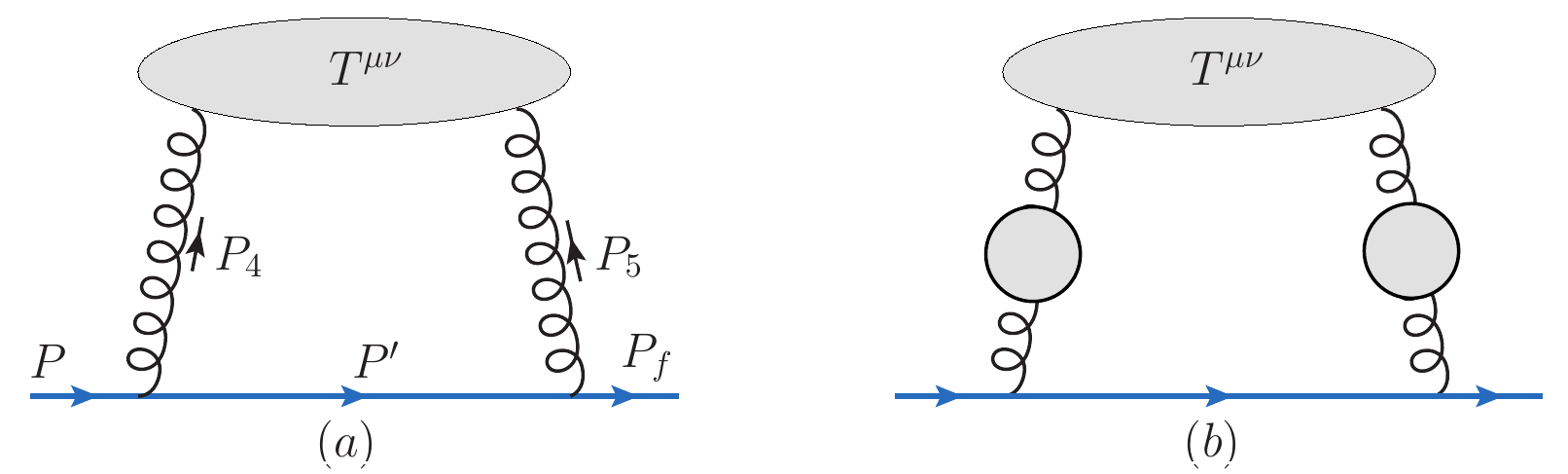, width=4in}
\caption{Feynman diagrams for the source term in QCD in the presence of fast parton with momentum $P$, which is represented by the blue fermion line. 
The diagram  (a) is at tree level and contributes to the hard region. The diagram in (b) contributes to the soft region with the hard 
thermal loop resummed gluon propagator is indicated by a gray blob.}
\label{fig-QCD}
\eef

Besides the external current approach that we have presented in the last subsection, 
the more general approach, which we adopt in this subsection, 
is to treat the energetic particle as a field rather than a classical current.  
This field then interacts with the medium through 
a two boson exchange and there is a resulting energy momentum transfer rate. We will 
take the case of a fast quark propagating through 
a QGP (with the massless quark/antiquark component only) as an example to illustrate 
the details of the calculation and  the resulting collisional energy loss. 
The setup is shown diagrammatically in Fig.~\ref{fig-QCD}, where the fast parton is represented by the blue fermion line. 
In Fig.~\ref{fig-QCD} we focus on the contribution from $F^{\tau\lambda}(P_4,P_5)$ in Eq.~(\ref{emtpiece}).
It includes an initial fast particle with momentum $P$, which interacts with the medium and scatters into final 
state $P_f$. We find it contributes
\ben
\label{sourcecont}
F^{\tau\lambda}(P_4,P_5)= -2 \pi i g^2 \int\frac{d^4 P'\,d^3\bald{P}_f}{P_f}\frac{P^\tau\,P'^\lambda 
+ P'^\tau\,P^\lambda - g^{\tau\lambda} P\cdot P'}{P'^2 + i\epsilon}\left[\delta^4(P-P_4-P')\delta^4(P'-P_5-P_f) 
+ P_4 \leftrightarrow P_5 \right]. \quad
\een
Combining Eq.~(\ref{sourcecont}) with Eq.~(\ref{emtpiece}) gives 
\ben
\label{finalemt}
\partial_\mu {T}^{\mu \nu}(X) &=& -8 i\,g^4N_fC_2 \int \frac{d^4 P_i}{(2\pi)^{11}} 
e^{-i X \cdot(P_4 + P_5)} T(P_3) D_{\sigma \tau}(P_4)D_{\omega \lambda}(P_5) G_A(P_3 - P_4)\nnu
&&\times\left[2\,({P_3}^\sigma\,{P_3}^\omega\,- {P_3}^\sigma\,{P_4}^\omega\,)P_5^{\nu} 
+ g^{\sigma\omega} \,P_3\cdot P_4\,P_5^{\nu} - g^{\nu\sigma}\,{P_3}^\omega(2 P_3\cdot P_5 - P_4\cdot P_5)\right] \nnu
&&\times\int\frac{d^4 P'\,d^3\bald{P}_f}{P_f}\frac{P^\tau\,P'^\lambda + P'^\tau\,P^\lambda - g^{\tau\lambda} P\cdot P'}
{P'^2 + i\epsilon}\left[\delta^4(P-P_4-P')\delta^4(P'-P_5-P_f) + P_4 \leftrightarrow P_5 \right].
\een
Eq.~(\ref{finalemt}) is then inserted in  Eq.~(\ref{eloss2}) to yield the collisional energy loss expression
\ben
\frac{dE}{dt}&=&8N_FC_2g^4\int\frac{d^4P_3d^4P_4}{(2\pi)^{7}}G_A(P_3-P_4)n_F(p_3)
\delta(P_3^2)D_{\sigma\tau}(P_4)D_{\omega\lambda}(-P_4)\frac{1}{E}\nnu
&&\times\left[2\,({P_3}^\sigma\,{P_3}^\omega\,- {P_3}^\sigma\,{P_4}^\omega\,)(-P_4)^{\nu} 
+ g^{\sigma\omega} \,P_3\cdot P_4\,(-P_4)^{\nu} - g^{\nu\sigma}\,{P_3}^\omega(-2 P_3\cdot P_4 + P_4^2)\right]\nnu
&&\times\left[\frac{P^{\tau}(P-P_4)^{\lambda}+P^{\lambda}(P-P_4)^{\tau}+g^{\tau\lambda}P\cdot P_4}{(P-P_4)^2+i\epsilon}
+\frac{P^{\tau}(P+P_4)^{\lambda}+P^{\lambda}(P+P_4)^{\tau}-g^{\tau\lambda}P\cdot P_4}{(P+P_4)^2+i\epsilon}\right].
\label{eq-dEdtdiag}
\een

In the soft region, we use the effective thermal propagator for the exchanged gluon, 
which is the same as that for the photon in Eq.(\ref{eq-propagator}) 
when the thermal photon mass is replaced with the thermal gluon mass. Using the same techniques as in the calculation of 
the collisional energy loss in the external current approach from the last section, we integrate over the angle of ${\bf p}_3$. 
The contribution from the interference between the longitudinal and transverse parts leads to zero, and the contributions 
from the longitudinal and transverse part are 
\ben
\left.\frac{dE}{dt}\right|_{LL}&=&4N_FC_2g^4\int\frac{dp_3dp_4dE_4}{(2\pi)^{3}}n_F(p_3)|\Delta_L(P_4)|^2{\rm sgn}(p_3-p_4\omega)p_3E_4^2,\\
\left.\frac{dE}{dt}\right|_{TT}
&=&2N_FC_2g^4\int\frac{dp_3dp_4dE_4}{(2\pi)^{3}}n_F(p_3)|\Delta_T(P_4)|^2{\rm sgn}(p_3-p_4\omega)
p_3E_4^2\left(1-\frac{E_4^2}{p_4^2}\right)^2. 
\een
The complete result in the soft region is the sum of the longitudinal and transverse parts and we enforce a upper 
limit cut-off $q^*$ of $\int dp_4$. We find  that the result matches the one from the previous  subsection 
in the limit of $u\to 1$ up to an additional color factor and the number of  quark flavors
\ben
\left.\frac{dE}{dt}\right|_{soft}&=&\frac{N_FC_2g^4T^2}{24\pi}\int_0^{q^*}dp_4\int_{-p_4}^{p_4}dE_4E_4^2
\left[|\Delta_L(P_4)|^2+\frac{1}{2}\left(1-\frac{E_4^2}{p_4^2}\right)^2|\Delta_T(P_4)|^2\right].
\een
In the hard region, we use the tree-level Feynman diagrams while ignoring any screening due to the plasma. 
The calculation is tedious but straightforward and yields
\ben
\left.\frac{dE}{dt}\right|_{hard}&=&-2N_FC_2g^4\int\frac{dp_3dp_4dP_4^0d\omega d\nu}{(2\pi)^{3}}p_3p_4^2n_F(p_3)
\left[\frac{1}{(p_4^0)^2-p_4^2}\right]^2\frac{p_4^0}{\sqrt{p_3^2+p_4^2-2p_3p_4\omega}}\nnu
&&\times\left\{\left[2p_3^2E(2+(1-\omega^2)+\nu^2(3\omega^2-1)-4\nu\omega)
-4p_3E(1-\omega\nu)(p_4^0-p_4\nu)-\left((p_4^0)^2-p_4^2\right)(p_4^0-p_4u)\right]\right.\nnu
&&\times\delta(p_4^0(p_3+E)-p_3p_4\omega-Ep_4\nu)\nnu
&&+\left[2p_3^2E(2+(1-\omega^2)+\nu^2(3\omega^2-1)-4\nu\omega)
-4p_3E(1-\omega\nu)(p_4^0-p_4\nu)+\left((p_4^0)^2-p_4^2\right)(p_4^0-p_4u)\right]\nnu
&&\left.\times\delta(p_4^0(p_3-E)-p_3p_4\omega+Ep_4\nu)\right\}.
\een
In the limit of $E\gg T\gg gT$, the integrals can be performed analytically
\ben
\left.\frac{dE}{dt}\right|_{hard}&=&\frac{N_FC_2g^4T^2}{48\pi}\left[\ln\frac{q_mT}{(q^*)^2}+\frac{8}{3}-12\ln(A)+\ln(4\pi)\right].
\een
Upon adding the hard and soft components the dependence on the separation scale $q^*$ cancels and the total collisional energy loss is 
\ben
\frac{dE}{dt}=\frac{N_FC_2g^4T^2}{48\pi}\left[\ln\frac{E}{g^2T}+2.725\right].
\een
One can immediately see that the collisional energy loss derived from the diagrammatic approach matches 
the one obtained  from the external current approach.

\section{Numerical result}
\label{sec-result}

In this section, we present the result for the  collisional energy and momentum loss of a fast lepton in an EPP 
 and of a fast quark in the QGP, respectively. For direct comparison, just like in the earlier sections, we only 
consider scattering of the external parton with the quark-antiquark component of the QGP.

In principle, in the case of strong interactions the method to match the hard and soft momentum exchange contributions 
by introducing the arbitrary intermediate momentum scale $gT\ll q^*\ll T$ is only valid in the weak coupling limit $g\ll 1$. 
Here, we extend the numerical evaluation to moderate values of $g$ for the purpose to investigating the dependence of 
collisional energy loss on the coupling constant $g$. The collisional energy loss as a function of the heavy quark energy 
$E$ and coupling $g$ is shown in Fig.~\ref{fig-dEdx-Eg}, where a constant temperature $T = 250 $ MeV and $N_F=3$ have been 
chosen as typical for  many phenomenological applications. The upper and lower surfaces are for charm quark and 
bottom quarks, respectively. One can see that, for both charm and bottom quarks, the magnitude of 
collisional energy loss in QGP increases monotonically with initial energy $E$ and coupling $g$. By comparing the collisional 
energy loss for charm and bottom quark, one clearly sees the large mass effect that goes $\sim \ln(E/M)$ in Eq.~(\ref{dEdt-QCD}) 
($\ln(1/\sqrt{1-u^2})=\ln(E/M)$). The energy loss of a bottom quark is approximately half of the energy loss of charm quarks
for $E \sim 10-15$~GeV. Note that one should not confuse the logarithmic growth of the collisional energy loss with larger 
jet quenching at higher energies $E$. It is the fractional energy loss    $\Delta E /E$ that enters phenomenological 
applications~\cite{Vitev:2005he,Kang:2012kc} and it goes $\sim \ln(E/M) / E$.
  
\bef
\psfig{file=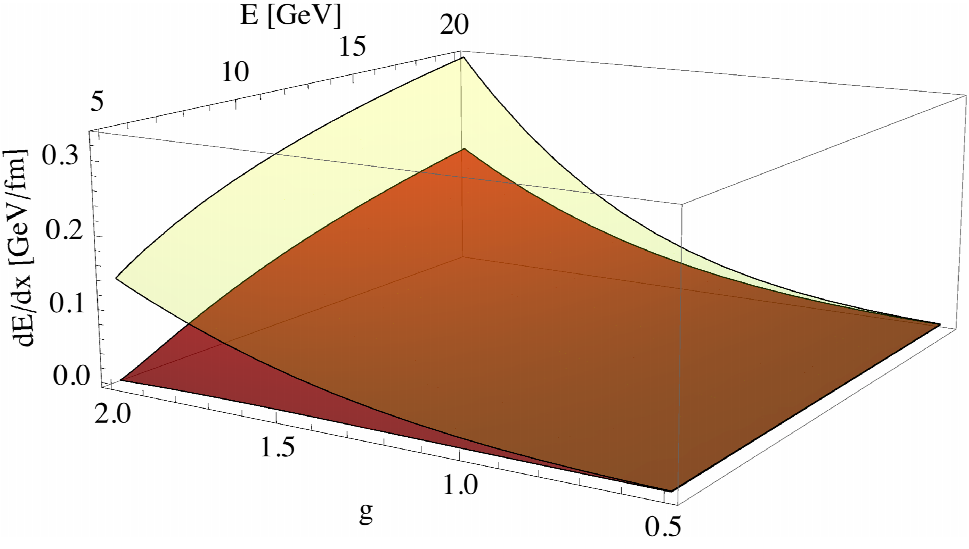, width=3.5in}
\caption{A three-dimensional representation of  collisional energy loss for charm (yellow surface) and bottom (red surface) 
versus energy $E$ and coupling $g$. We have chosen  $M_c=1.5$ GeV and $M_b=4.5$ GeV, respectively,  a constant temperature 
$T = 250 $ MeV and $N_F=3$ as typical  for many phenomenological applications of heavy ion collisions.}
\label{fig-dEdx-Eg}
\eef

We also show in Fig.~\ref{fig-dEdx-Tu} the collisional energy loss (left panel) and momentum loss (right panel), where 
a fixed coupling $g=1$ has been chosen. One can see that the collisional momentum loss behaves similarly to the energy loss, 
both of them increase monotonically with the temperature $T$ and velocity $u$.   
In the small and moderate velocity regimes,  the momentum loss is larger than the energy loss, due to the finite 
velocity effect, as one can see from Eq. (\ref{eq-dpdt}). 
We mention here that in the ultrarelativistic limit $u\to 1$ and nonrelativistic limit $u\to 0$, the formula Eq. (\ref{dEdt-QCD}) 
for collisional energy loss that we have used for the numerical evaluation of Fig. \ref{fig-dEdx-Tu} breaks down. Therefore, 
physical kinematics should be imposed to compute the collisional energy as we presented in the last section.
\bef
\psfig{file=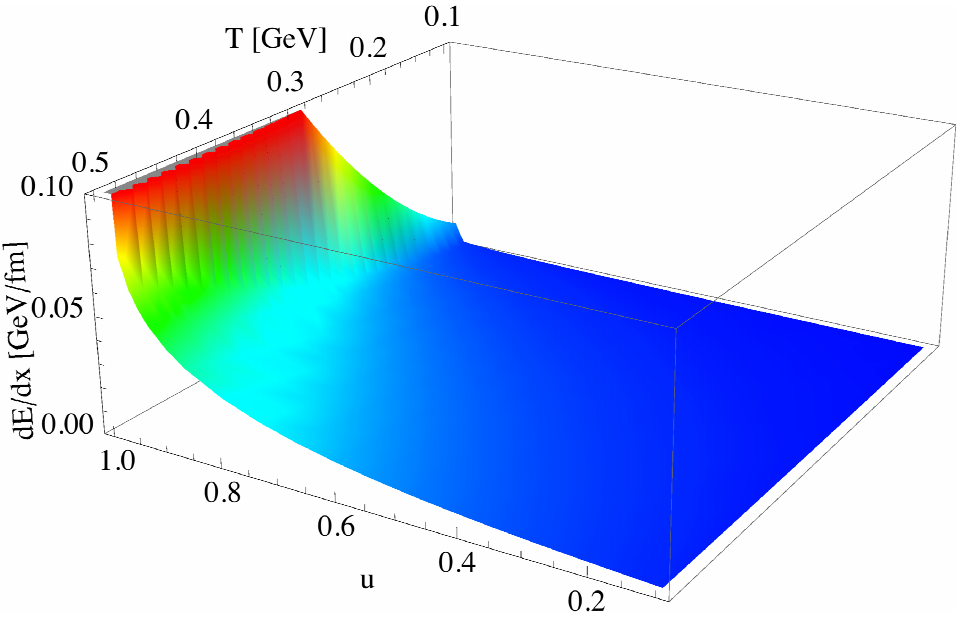, width=3.5in}
\psfig{file=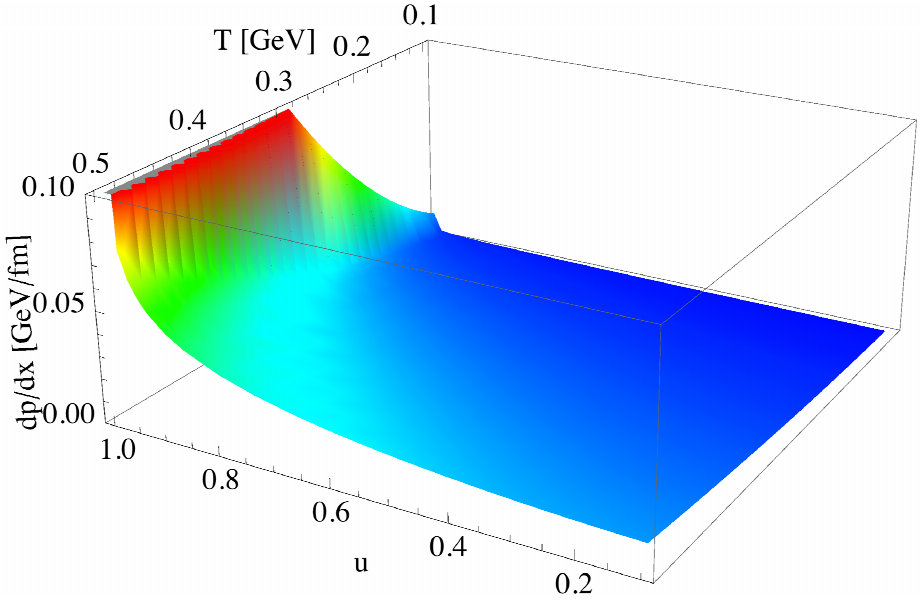, width=3.5in}
\caption{Three-dimensional representations of collisional energy loss (left) and momentum loss (right) versus 
the temperature $T$ and  velocity $u$. A fixed coupling $g=1$ has been chosen for the numerical evaluation. One 
can see that the   momentum loss is larger than the energy loss in the small and moderate  velocity regions.}
\label{fig-dEdx-Tu}
\eef

\bef
\psfig{file=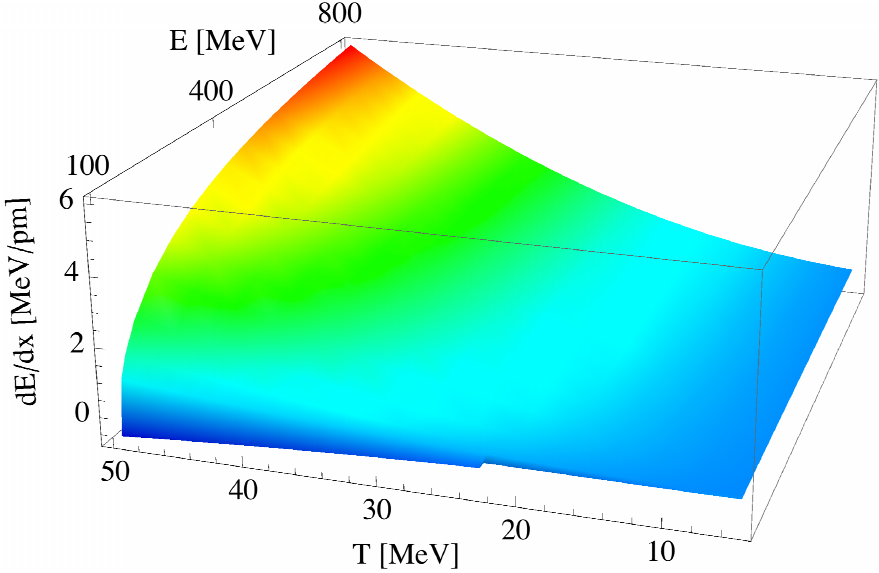, width=3.5in}
\caption{A three-dimensional representation of the collisional energy loss  versus the energy $E$ and temperature $T$ for 
a heavy muon propagating through an EPP. The muon mass is $M_{\mu}=105$ MeV, the elementary charge  is 
fixed at $e=0.3$ (in terms of natural units), corresponding to a fine structure constant $\alpha=1/137$.}
\label{fig-dEdx-ET}
\eef
Next, we extend the numerical examples to the case of  
 of relativistic QED plasmas that can be produced in high-intensity laser fields and play a role in various astrophysical 
situations, such as in supernova explosions. We show in Fig.~(\ref{fig-dEdx-ET}) 
 an example of the collisional energy loss from temperatures $T=10 - 50 $~MeV that can be typically realized in laser produced 
and supernovae electron-position palsmas~\cite{Thoma:2008my}. In the relativistic plasma we considered here,  $T\gg m_e$, 
so that the mass of the electron and positron in the medium can be neglected. The elementary charge $e=0.3$, 
corresponding to a fine structure constant $\alpha=1/137$,  indicates that one can distinguish the soft and hard momentum scales, 
i.e. $eT\ll T$, and the EPP is weakly coupled.  Therefore, the result we derived in this paper should give a good description of 
the physics. Shown in Fig.~(\ref{fig-dEdx-ET}) is the collisional energy loss in QED as functions of the initial muon energy $E$ and medium 
temperature $T$, the collisional energy loss increases with the increasing of the medium temperature, for example, this leads 
an energy loss of $3$ MeV/pm for a muon with energy $E=200$ MeV at $T=50$ MeV.

\section{Summary}
\label{sec-summary}
In this paper we considered energetic charged particle  propagation in an EPP and a QGP (quark-quark scattering only). We derived 
the  energy and momentum  absorbed by the medium per unit time when the particle is slowed down due to collisional interactions. 
For this purpose, starting from the medium's point of view, we provided an operator definition of collisional energy and momentum 
transfer rate based upon the divergence of the medium EMT. By using an external current approach, we evaluated the energy 
and momentum loss of an energetic lepton passing through a thermal electron-positron plasma. Furthermore, in a more general
diagrammatic approach we considered  the collisional energy loss of a fast parton in the QGP.
In both cases we applied the method used by BT to separate the exchanged momenta into hard and soft regions, 
to evaluate the relevant HTL resumed and tree level diagrams.  We showed explicitly that the newly 
developed formalism leads to  results which are infrared safe, independent of gauge and any 
separation scales. Our results for the energy-momentum absorption rate by the medium reproduce (up to the anticipated minus sign) 
the well known results  for the energy loss of energetic charged particle from the scattering rate approach.
To illustrate the analytic results, we gave examples of an energetic heavy quark propagating through the quark-gluon plasma, 
which can be produced  in heavy-ion collisions, and an energetic muon traveling in electron-positron plasma, which can be produced 
in high-intensity laser fields. In summary, we found that up to the expected difference in the energy, temperature, coupling strength
and degrees of freedom,  the collisional energy-momentum transfer rates in QED and QCD behave very similarly. It will be instructive 
in the future to carry out such comparison beyond the weakly coupled regime using numerical techniques.
\\

\small{{\it Acknowledgments}:  This work was supported in part by the US Department of Energy, Office of Science.}

\end{document}